\documentclass[pre,twocolumn]{revtex4-2}
\usepackage{graphicx}
\usepackage{amsopn}
\usepackage{amsfonts}
\usepackage{amsmath}
\usepackage{latexsym}
\usepackage{bm}
\usepackage{amssymb}
\usepackage{color}

\begin{document}

\title{Correlations of the phase gradients of the light wave \\
propagating in a turbulent medium in the regime of weak scintillations}

\author{V.A.Bogachev$^{1}$,  I.V. Kolokolov$^{2}$, V.V. Lebedev$^{2}$, A.V.Nemtseva$^{1,3}$ and F.A.Starikov$^{1,4}$}

\affiliation{
$^1$Russian Federal Nuclear Center -- Russian Research Institute of Experimental Physics \\
37 Mira ave, Sarov, Nizhny Novgorod region, 607188, Russia;  \\
$^2$ Landau Institute for Theoretical Physics, RAS, \\
142432, Chernogolovka, Semenova 1A, Moscow region, Russia; \\
$^3$Lomonosov State University, Faculty of Physics,  \\
Branch of Lomonosov State University, 2 Parkovaya St., Sarov, Nizhny Novgorod Reg., 607328 Russia \\
$^4$National Research Nuclear University MEPhI -- Sarov Physical and Technical Institute, \\
6 Dukhova ave, Sarov, Nizhny Novgorod region, 607186, Russia. }

\date{\today}

\begin{abstract}

We investigate numerically correlation functions of the phase of light waves that propagate through turbulent media. Special attention is paid to the off-diagonal component of the correlation function of the phase gradients which is insensitive to the outer scale of turbulence. The results of our simulations are in a good agreement with the analytical expressions obtained in Ref. \cite{KLS25}. Thus, we numerically confirm expectations based on uniformity of the perturbation theory for the logarithm of the envelope.

\end{abstract}


\maketitle

\section{Introduction}
\label{sec:intoduction}

This study focuses on numerical simulations of light beam propagation through a turbulent medium, such as the Earth's atmosphere. Turbulence causes fluctuations in the refractive index of the medium, leading to distortions of the light beam. Observations and theoretical arguments have revealed that the character of these distortions depends on the path taken by the beam. For relatively short distances, the distortions are minor, and this situation is known as the weak scintillation regime. However, over longer distances, the light beam splits into individual speckles, this case is known as the strong scintillation regime.

The main ideas exploited for understanding peculiarities of light propagation in a turbulent medium can be traced back to the classic works of Kolmogorov and Obukhov \cite{Kolm411,Kolm412,Obukhov41} based on the concept of the energy cascade. In the works scaling laws are established characterizing fluctuations of the velocity within the so-called inertial interval of scales of turbulence. Subsequently, Obukhov \cite{Obuk49} and Corrsin \cite{Corr51} expanded this approach to include fluctuations of a passive scalar, such as temperature or impurity density, in a turbulent medium. These findings can be applied to describe the statistical properties of fluctuations in the refractive index as well.

The problems related to the light propagation in a turbulent medium were intensively studied in the second half of the twentieth century. The results of the investigations are outlined in Refs. \cite{Tat67,Goodman,Tat75,Tat751,StrohbehnBook1978,AP98}. They concern mainly the weak scintillation regime. Besides, some results for the strong scintillation regime were also obtained, see, e.g., Ref. \cite{Char94}. Recently, interest in the problem has been renewed, mainly due to the increasing capabilities of numerical simulations, which allow us to obtain detailed information about light propagation in a turbulent medium, see Refs. \cite{Vorontsov2010,Vorontsov2011,Lachinova2016,Lushnikov2018,Aks19}.

Distortions caused by atmospheric turbulence can be partially corrected using a technique called adaptive optics, see, e.g., Refs. \cite{Beck93,Tys10}. In Ref. \cite{BKLS24}, we have proposed a new method for measuring the Fried parameter, $r_0$, see Ref. \cite{Fried75}. The parameter is a crucial characteristic of radiation in the turbulent atmosphere determining the requirements for spatial resolution in adaptive optics systems. The idea of Ref. \cite{BKLS24} is in exploiting the off-diagonal component of the phase gradient correlation function, which can be measured, say, using a Shack-Hartmann wavefront sensor \cite{Shack01}. The expressions presented in Ref. \cite{BKLS24} were derived for the case where the separation $r$ between the observation points is much larger than the radius of the first Fresnel zone. The results reported in Ref. \cite{KLS25} expand the approach to arbitrary separations $r$.

The explicit expressions presented in Refs. \cite{BKLS24} and \cite{KLS25} were derived from the linearized equation for the logarithm of the envelope of the wave packet. A necessary condition for the application of this approach is the smallness of the Rytov variance $\sigma_R^2$ \cite{Tat67}. However, Refs. \cite{Varva,Baraban,Tat751} state that the inequality $\sigma_R^2 \ll 1$ may not be sufficient to accurately calculate the correlation function of the phase for all distances $r$ using perturbation theory. The reason for this is that there may be terms in the perturbation series that are singular at $r\to 0$ or $r\to \infty$, which violates the uniformity of the approximation. Indeed, the expansion in $\sigma_R^2$ of the correlation functions of the envelope $\Psi$ ceases to be valid for sufficiently large $r$.

However, in the perturbation series for the correlation functions of the phase of the envelope, only gradients of the logarithm of the envelope are present, see Subsec. \ref{subsec:log}. That makes the perturbation theory for the phase uniform in $\sigma_R^2$. To confirm this assertion, we have calculated the next order in $\sigma_R^2$ for the phase correlation function, which demonstrates explicitly the uniformity of the expansion. A complete analytical scheme that allows us to estimate the highest orders of the perturbation theory series will be published elsewhere \cite{KL26}.

In this paper, we use direct numerical modeling to show the applicability of the leading approximation in $\sigma_R^2\leq 1$, derived in Ref. \cite{KLS25} over a wide range of separations $r$. Our simulations were performed using the traditional approach \cite{Kan96}, in which a continuous turbulent medium is modeled using a series of screens, on which random phase shifts of the light wave occur. The wave propagates freely between the screens. The statistics of these phase shifts is chosen to mimic the effect of continuous fluctuations in the refractive index on the light wave.

The structure of our paper is as follows. Section \ref{sec:mainrel} contains the basic relations concerning the light propagation in a turbulent medium, including fluctuations of the refractive index, the envelope of the wave packet and its logarithm. Here we recall shortly our previous findings. Section \ref{sec:numerics} is devoted to describing results of our numerical simulations conducted for the initial plane wave. Here we compare our numerical results to analytics. In Section \ref{sec:conclusion}, we summarize our findings and present the main conclusions of our research.

\section{Main relations}
\label{sec:mainrel}

We consider light wave packets propagating in a turbulent medium, such as the atmosphere, assuming that we are dealing with a monochromatic wave. These wave packets can be characterized by their envelope, $\Psi$, which describes the structure of the wave packet on scales larger than its wavelength. The envelope $\Psi$ is a complex field determining, say, the electric field, which is proportional to
\begin{equation}
\mathrm{Re}\, [\Psi \exp(ik_0z)],
\nonumber
\end{equation}
where $k_0$ is the wave vector of the carrying wave and $z$ is the coordinate in the direction of the wave packet propagation.

The envelope $\Psi$ is controlled by the following parabolic equation
\begin{equation}
i \partial_z \Psi +\frac{1}{2k_0}\nabla^2 \Psi
+k_0 \nu  \Psi=0.
\label{gain1}
\end{equation}
Here $\nu$ designates a fluctuation of the refraction index and the vector operator $\nabla$ in Eq. (\ref{gain1}) designates the coordinate gradient in the direction perpendicular to the $Z$-axis, $\nabla=(\partial_x,\partial_y)$. The equation (\ref{gain1}) implies that the envelope, $\Psi$, adapts simultaneously to the state of the medium, which is described by the field $\nu$. This property is justified by the speed of light being very high.

A theoretical investigation of light propagation based on the equation (\ref{gain1}) involves, generally, several steps. First of all, one introduces a source generating some initial envelope $\Psi_0(x,y,0)$ at $z=0$. Then one solves the evolution equation (\ref{gain1}) along $z$ with the initial condition $\Psi_0(x,y,0)$ at a given field $\nu$. And, finally, one finds correlation functions of $\Psi,\Psi^\star$ (where star designates complex conjugation) at the final $z$, where a receiver is placed. The correlation functions, found by averaging over realizations of $\nu$, contain the complete information about statistical properties of the light beam at the receiver.

Of course, it is impossible to conduct the program explicitly. That is why some approximate methods are used to evaluate the correlation functions. One such method is the perturbation theory used in the regime of weak scintillations. In the following, we discuss the conditions under which the perturbation theory can be applied. Let us first introduce some quantities that will be needed for this discussion.

Statistical properties of the refraction index $\nu$ can be characterized by its second order structure function. In the inertial range of turbulence the structure function follows a power law
\begin{eqnarray}
\langle [\nu  (\bm r_1,z_1) -\nu  (\bm r_2,z_2)]^2\rangle
=C_n^2 (r^2+z^2)^{1/3},
\label{KolmOb}
\end{eqnarray}
characteristic of a passive scalar, see Refs. \cite{Obuk49,Corr51}. Here and below the angular brackets denote time averaging. In the expression (\ref{KolmOb}), the quantity $C_n^2$ determines the strength of the fluctuations of the refractive index, $\bm r=\bm r_1-\bm r_2$, $z=z_1-z_2$. The expression (\ref{KolmOb}) suggests that the turbulent fluctuations in the inertial range are statistically homogeneous and isotropic, which is consistent with experimental observations \cite{Monin,Frisch}.

We assume that the propagation distance of the wave packet is much larger than all its characteristic lateral scales. Then, the field $\nu$ can be treated as a quantity which is short correlated along the $Z$-axis. The property motivates the approximation
\begin{eqnarray}
\langle \nu  (\bm r_1,z) \nu  (\bm r_2,\bm z_1)\rangle=
\delta(z-z_1) C_n^2 {\mathcal A}(\bm r),
\label{delta}
\end{eqnarray}
where $\bm r=\bm r_1-\bm r_2$. The factor $C_n^2$ in Eq. (\ref{delta}) can be thought as a gradually varying function of $z_1$. The characteristic scale of the variations should be larger than the lateral scales.

The function $\mathcal A$ in Eq. (\ref{delta}) is determined by the spectrum of the fluctuations of the refractive index. The von Karman spectrum \cite{Kar48} $\propto (k^2+\kappa^2)^{-5/6}$ is widely used, where $k$ is wave vector and $\kappa$ determines the outer scale of turbulence $L_0$, $L_0=2\pi/\kappa$. For the von Karman spectrum \cite{BKLS24}
\begin{equation}
{\mathcal A}(\bm r)=
{\mathcal A}_0  -1.4572\, r^{5/3}
+ 1.1727 \kappa^{1/3} r^2,
\label{corrfr2}
\end{equation}
where ${\mathcal A}_0\sim L_0^{5/3}$. The expression (\ref{corrfr2}) is derived under the assumption that $r$ is much smaller than $L_0$ and represents the leading order terms in the small parameter $r/L_0$.

The correlation length of the envelope $\Psi$ in the transverse direction, the Fried parameter $r_0$, is defined in accordance with Ref. \cite{Fried75}
\begin{equation}
r_0^{-5/3}= 0.423 \, k_0^2 \int_0^z d\zeta\, C_n^2(\zeta).
\label{Fried}
\end{equation}
For the homogeneous medium where $C_n^2$ is independent of $z$
\begin{equation}
r_0^{-5/3}= 0.423 \, k_0^2 z C_n^2.
\label{Fried2}
\end{equation}
The value of $r_0$ diminishes as the distance $z$ travelled by the wave packet grows.

Traditionally, the level of scintillations is characterized by a dimensionless parameter called the Rytov variance $\sigma_R^2$. For a perturbed plane wave the Rytov variance is defined as
\begin{equation}
\sigma_R^2=2.25 \, k_0^{7/6} z^{5/6}
\int_0^z d\zeta\, (1-\zeta/z)^{5/6} C_n^2(\zeta),
\label{pla2}
\end{equation}
see, e.g., Ref. \cite{Spencer2021}. For the homogeneous medium one finds
\begin{equation}
\sigma_R^2=1.23\,  C_n^2 k_0^{7/6} z^{11/6},
\label{rytovd}
\end{equation}
after taking the integral in Eq. (\ref{pla2}). The quantity $\sigma_R^2$ grows as the wave packet runs.

It is possible to develop a perturbation theory for correlation functions of $\Psi,\Psi^\star$. As it was noted in Ref. \cite{Tat67} the applicability condition for perturbation theory is related to smallness of the Rytov variance $\sigma_R^2$ (\ref{rytovd}). However, the terms of the perturbation series grow as the distance between the points increases. Therefore, to obtain reasonable results a deep resummation of the perturbation series is needed \cite{Char94,KLL20}. To avoid the difficulty we pass to the equation for the logarithm of the envelope.

\subsection{Logarithm of the envelope}
\label{subsec:log}

Let us pass from the envelope $\Psi$ to its logarithm $\ln \Psi$. The phase $\varphi$ of the envelope $\Psi$ is the imaginary part of $\ln \Psi$, whereas the real part of $\ln \Psi$ determines the absolute value $|\Psi|$ of the envelop. The equation (\ref{gain1}) leads to the following equation for the logarithm of $\Psi$
\begin{equation}
\partial_z \ln\Psi
-\frac{i}{2k_0}\nabla^2 \ln\Psi
-\frac{i}{2k_0}(\nabla \ln\Psi)^2
-i k_0 \nu =0.
\label{logpsi}
\end{equation}
Note that, unlike the linear equation (\ref{gain1}), the equation (\ref{logpsi}) is nonlinear, it has a second-order nonlinearity.

The logarithm of the envelope, $\ln \Psi$, includes both the contribution from the free propagation of the envelope (without random diffraction) and the fluctuation contribution due to $\nu$. It is convenient to exclude the first contribution passing to the fluctuating quantity $\eta=\ln (\Psi/\Psi_0)$, where $\Psi_0$ is the envelope for the free propagation of the wave. The envelope $\Psi_0$ is a solution of the equation (\ref{logpsi}) for $\nu =0$, depending on the initial envelope profile. The equation for the quantity $\eta$ can be derived from the equation (\ref{logpsi}), it is
\begin{eqnarray}
\partial_z \eta
-\frac{i}{2k_0}\nabla^2 \eta
-\frac{i}{k_0} \nabla \ln\Psi_0 \nabla\eta
\nonumber \\
=i k_0 \nu +\frac{i}{2k_0}(\nabla \eta)^2.
\label{log1}
\end{eqnarray}
Note that the equation is, generally, spatially inhomogeneous due to the presence of the term containing $\Psi_0$. The equation (\ref{log1}) has to be solved with the zero initial condition $\eta=0$ at $z=0$. Taking the equation complex conjugated to Eq. (\ref{log1}) one finds the equation for $\eta^\star$.

One introduces the pair correlation functions
\begin{eqnarray}
\varPhi(\bm r_1,z_1;\bm r_2,z_2)
=\langle \eta(\bm r_1,z_1)
\eta(\bm r_2,z_2) \rangle,
\label{pairphi} \\
\varXi(\bm r_1,z_1;\bm r_2,z_2)
=\langle \eta(\bm r_1,z_1)
\eta^\star(\bm r_2,z_2) \rangle,
\label{pairxi}
\end{eqnarray}
related to $\eta$. Here, angular brackets denote time averaging, or, in a theoretical context, averaging over the statistics of $\nu$. The pair correlation function of the phase fluctuations $\varphi$ of the envelope $\Psi$ is expressed as
\begin{equation}
\langle \varphi(\bm r_1,z_1) \varphi(\bm r_2,z_2) \rangle
=-\frac{1}{4}\left(\varPhi+\varPhi^\star-\varXi -\varXi^\star\right),
\label{log3}
\end{equation}
where all the functions on the right-hand side of the equation have the arguments $r_1, z_1; r_2, z_2$.

The correlation functions (\ref{pairphi},\ref{pairxi}) can be found if one neglects the nonlinear term in the equation (\ref{log1}). Then the equation (\ref{log1}) becomes linear and $\eta$ can be expressed via $\nu$ using the Green's function. Then one finds the correlation functions (\ref{pairphi},\ref{pairxi}) averaging the products $\eta_1 \eta_2$ and $\eta_1 \eta_2^\star$ by using Eq. (\ref{delta}) for different $\Psi_0$.

The program was conducted in Ref. \cite{KLS25} for the initial profiles $\Psi_0(x,y,0)$ corresponding to a plane wave and to a point source. Then the correlation functions (\ref{pairphi},\ref{pairxi}) are written in terms of the confluent hypergeometric function. Passing to the pair correlation function of the phase (\ref{log3}) and taking the gradients one finds the pair correlation function of the phase derivatives
\begin{equation}
Q_{\alpha\beta}=
\left\langle \frac{\partial}{\partial r_{1\alpha}}\varphi(\bm r_1,z)
\frac{\partial}{\partial r_{2\beta}}
\varphi(\bm r_2,z) \right\rangle.
\label{qalbe}
\end{equation}
It is written explicitly for two noted cases, see Ref. \cite{KLS25}.

For the initial plane wave the pair correlation function of the phase derivatives is written as
\begin{eqnarray}
Q_{\alpha\beta}^{(0)}=1.435 \frac{1}{r_0^2}\left(\frac{r_0}{r}\right)^{1/3}
\left\{\frac{5}{3}\left[1+f_1\left(\xi\right)\right]
\delta_{\alpha\beta}\right.
\nonumber \\ \left.
+\frac{1}{3}\left[1+f_2\left(\xi\right)\right]
\left( \delta_{\alpha\beta}
-2\frac{r_\alpha r_\beta}{r^{2}}\right)\right\}
-\frac{10.2}{L_0^{1/3} r_0^{5/3}}\delta_{\alpha\beta}
, \quad
\label{zeroapp}
\end{eqnarray}
where
\begin{equation}
\xi=k_0 r^2/(4z) ,\qquad
\label{xixixi}
\end{equation}
and the functions $f_1(\xi),f_2(\xi)$ are zero for $\xi=0$ and tend to unities as $\xi\to\infty$. The plots of the functions can be found in Ref. \cite{KLS25}. An interesting property of the expression (\ref{zeroapp}) is that { the off-diagonal quantity $Q_{xy}$ is independent of $L_0$ and reaches a local maximum at}
\begin{equation}
r^\ast =1.66 (\sigma_R^2)^{3/5} r_0.
\label{rstar}
\end{equation}
The property allows one to extract an additional information about the Rytov variance from experimental data.

{
Corrections to the pair correlation functions (\ref{zeroapp}) are related to the nonlinear term in Eq. (\ref{log1}). The first non-vanishing corrections appear in the second order of the expansion in the non-linearity. It is possible to analyze the first correction in detail. The results of the analysis will be published elsewhere \cite{KL26}. Omitting technical details, we present here only the main results.

At small distances, where $r \ll \sqrt{z / k_0}$, there appear singular contributions in perturbation theory. However, these contributions cancel each other out when calculating $Q^{(1)}_{\alpha \beta}$. As a result, for $r\lesssim \sqrt{z/k_0}$ the following estimation is valid:
\begin{equation}
Q^{(1)}_{\alpha\beta}/Q^{(0)}_{\alpha\beta}\sim \sigma_R^2.
\label{sr}
\end{equation}
Note that this estimate is valid despite the fact that both, $Q^{(0)}_{\alpha\beta}$ and $Q^{(1)}_{\alpha\beta}$ diverge as $r\to 0$, see Eq. (\ref{zeroapp}).

For large separations, $r\gg \sqrt{z/k_0}$, the correction to $Q^{(0)}_{\alpha\beta}$ takes an additional smallness:
\begin{equation}
Q^{(1)}_{\alpha\beta}/Q^{(0)}_{\alpha\beta}\sim \sigma_r^2\frac{z}{k_0 r^2}\ll \sigma_R^2.
\label{lr}
\end{equation}
This estimation shows that the applicability of perturbation theory is improving, despite the increase of the  phase fluctuations. This is due to the fact that the nonlinear term in the equation (\ref{log1}) contains spatial derivatives of the field $\eta$, rather than its amplitude. In their turn, correlations of gradients $\nabla \eta$ decrease with increasing $r$.

It is possible to verify that such inequalities take place in higher orders of perturbation theory as well. This means the uniformity of the approximation for $Q_{\alpha\beta}$ based on expansion in $\sigma_R^2$.} Thus, one expects that the expression (\ref{zeroapp}) describes well observable data at the condition $\sigma_R^2\ll1$. The main focus of this work is to check the assertion numerically.

\section{Numerics}
\label{sec:numerics}

Here we present results of our numerical simulations. Our setup corresponds to a monochromatic wave with a wavelength of $\lambda = 0.55 \mu m$, which propagates along a statistically homogeneous path. The initial profile has a width of $1.5$ meters and is characterized by a flat central part and super-Gaussian wings. The width is much greater than the size of the inhomogeneities associated with fluctuations in the refractive index. Thus, in the region near the axis of the beam the regime of a plane wave is realized. The beam path is characterized by specific values of the Fried parameter $r_0$ and of the Rytov dispersion $\sigma_R^2$.

The beam propagation is assumed to be governed by the parabolic equation (\ref{gain1}). Numerically, the equation (\ref{gain1}) was solved by the finite difference method with splitting diffraction and refraction processes, see, e.g., Ref. \cite{Fle76}. The continuous turbulent medium is modeled by a chain of infinitely thin phase screens that are perpendicular to the direction of beam propagation. At each screen, a random phase jump of the envelope is generated, see, e.g., Ref. \cite{Kan96}. The computation process involves a series of consecutive transformations of the envelope $\Psi$, including the addition of the phase jumps at the screens and the free diffraction of $\Psi$ between them.

The diffraction of radiation in the regions between the screens was calculated by solving the propagation equation using the Ladagin \cite{Lad85} difference scheme, which has zero amplitude error and a fourth-order phase error when integrating the diffraction operator. A chain of the screens was chosen in such a way that, at the evolution between the screens, the phase shift due to diffraction was small, and the phase dispersion on the screen did not exceed $1$ rad$^2$. The smallest scale of spatial inhomogeneities, that is, the effective internal scale of turbulence, in the numerical model is determined by the step of the computational grid and lies in the range of $0.25\div 1$ cm.

The addition of the phase jump $\Delta\varphi$ at a screen means the following transformation of the envelope
\begin{equation}
\Psi_+(x,y)=\exp[i\Delta\varphi(x,y)]\Psi_-(x,y).
\label{transf}
\end{equation}
Here $\Psi_-$ is the envelope before the screen and $\Psi_+$ is its value after the screen. Note that the transformation (\ref{transf}) does not touch the absolute value of the envelope $\Psi$.

The phase jumps for different screens are independent random functions. For each screen it is generated using the pair correlation function
\begin{equation}
\langle \Delta\varphi(\bm r_1) \Delta\varphi(\bm r_2) \rangle
=   C_n^2 k_0^2 {\mathcal A}(\bm r) \Delta z,
\label{screen}
\end{equation}
where $\bm r=\bm r_1-\bm r_2$ and $\Delta z$ is the separation between the screens. The expression (\ref{screen}) represents the integral influence of the random diffraction index $\nu$ in the layer of width $\Delta z$ in accordance with Eqs. (\ref{delta},\ref{log1}). The function ${\mathcal A}(r)$ was chosen to be formed by the von Karman spectrum \cite{Kar48}
\begin{equation}
{\mathcal A}(r)=
\int \frac{d^2 q}{(2\pi)^2}
\exp(i\bm q \bm r) \frac{8.19}{(q^2+\kappa^2)^{11/3}},
\label{screen2}
\end{equation}
where $\kappa=2\pi/L_0$.

The generation of random phase jumps $\Delta\varphi$ on the screens was carried out using a combination of the usual spectral method \cite{Kan96} and the method of subharmonics \cite{Kan98,Joh94} (for details see Ref. \cite{Bog24}). In the spectral method, a random phase field is formed by filtering a Gaussian pseudorandom field, and the transfer function of the filter is determined by the spatial spectrum of phase fluctuations. This method allows one to form phase screens with the external scale of turbulence $L_0 \leq D/2$, where $D$ is the transverse size of the counting area.

The subharmonic method is used to simulate large-scale distortions with $L_0>D/2$ while maintaining the size of the counting area $D$. The main idea behind the subharmonic method is to densify the nodes of the computational grid in the spectral plane around the zero harmonic. It is done iteratively by adding harmonics (subharmonics) to the low-frequency part of the spatial spectrum. At each iteration, additional $32$ harmonics are added to the phase spectrum with increments three times smaller than the previous one. Thus, the accuracy of reproducing the low-frequency part of the phase spectrum increases with increasing number of iterations. In our computations, the number of iterations ranged from $4$ to $8$.

At the end of the turbulent route, a profile of the envelope $\Psi$ was recorded. During the repeated passage of the wave along random implementations of the turbulent route by the Monte Carlo method, a random set of $\Psi$ was formed. In the calculations, the size of the set ranged from $300$ to $500$. Using the set, one can find averaged characteristics of the radiation.

\begin{figure}
\begin{center}
\includegraphics[width=0.5\textwidth]{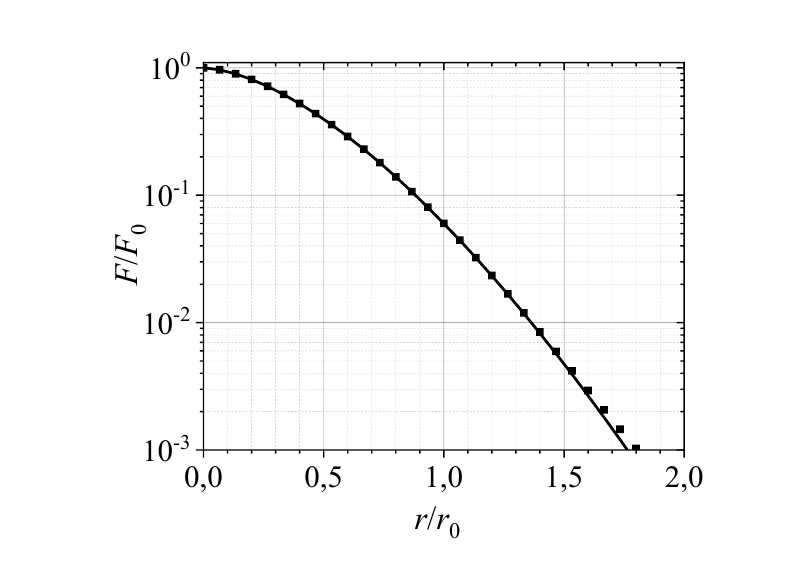}
\end{center}
\caption{The pair correlation function (\ref{pairnum}) as a function of $r/r_0$ for $\sigma_R^2=0.5$, $r_0=3.75$ cm, $L_0=20$ m. The solid line is drawn using the analytic expression (\ref{BKLS24}) and the filled squares represent the numerical data extracted from the simulations.}
\label{fig:Fig1}
\end{figure}

The accuracy of the calculated model was determined as follows. For a set of transverse distributions $\Psi(x.y)$, recorded at the end of the path, the pair correlation function was calculated
\begin{equation}
F_{num}(\bm r_1,\bm r_2)=
\langle \Psi(\bm r_1) \Psi^\star(\bm r_2) \rangle,
\label{pairnum}
\end{equation}
where $\bm r_1=(x_1,y_1)$, $\bm r_2=(x_2,y_2)$, and the angular brackets designate averaging over the set. For statistically homogeneous and isotropic turbulence, the pair correlation function depends only on the distance between the points, $\bm r=\bm r_1-\bm r_2$, and is real. The properties are revealed after averaging, indeed.

For the initial plane wave passing the turbulent path, where fluctuations of the refractive index are determined by the von Karman spectrum, the normalized pair correlation function is
\begin{eqnarray}
\ln \left[F(r)/F(0)\right] =
- 3.44  \left(\frac{r}{r_0}\right)^{5/3}
\nonumber \\ \times
\left[1-1.485 \left(\frac{r}{L_0}\right)^{1/3}\right].
\label{BKLS24}
\end{eqnarray}
The expression was obtained analytically at small $r/L_0$, see \cite{Shis68,Tat69}.

A comparison of the analytical expression (\ref{BKLS24}) and the numerical data obtained from simulations for specific values of the parameters $\sigma_R^2$, $r_0$ and $L_0$ is shown in Fig. \ref{fig:Fig1}. It can be seen that the numerical data reproduce the analytics pretty well, which indicates that all the main turbulence parameters are adequately reproduced in our simulations.

\begin{figure}
\begin{center}
\includegraphics[width=0.5\textwidth]{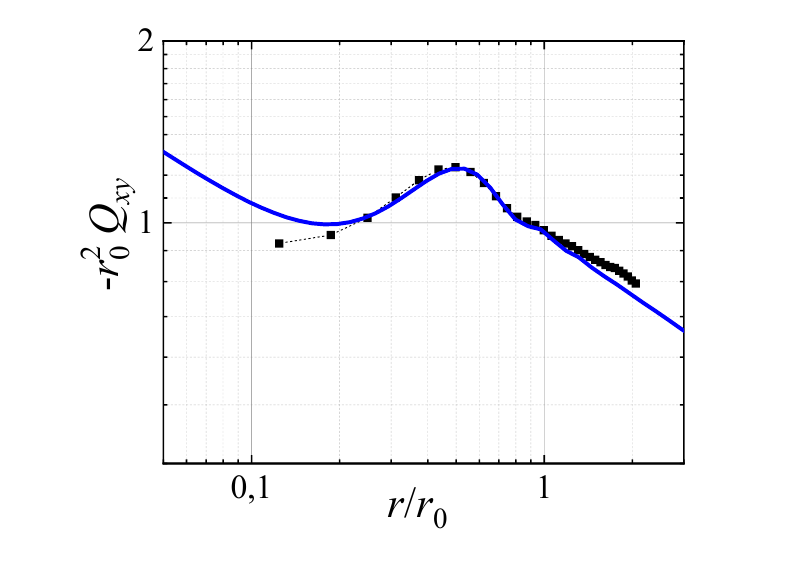}
\end{center}
\caption{The product $-r_0^2 Q_{xy}$ as a function of $r/r_0$ taken at $x=y=r/\sqrt2$ for $\sigma_R^2=0.14$, $r_0$=5.69 cm, $L_0$=20 m, where $Q_{xy}$ is the non-diagonal component of the pair correlation function of the phase gradients. The solid line is drawn using the analytic expression given in Ref. \cite{KLS25} and the filled squares represent the numerical data extracted from the simulations.}
\label{fig:Fig2}
\end{figure}

\begin{figure}
\begin{center}
\includegraphics[width=0.5\textwidth]{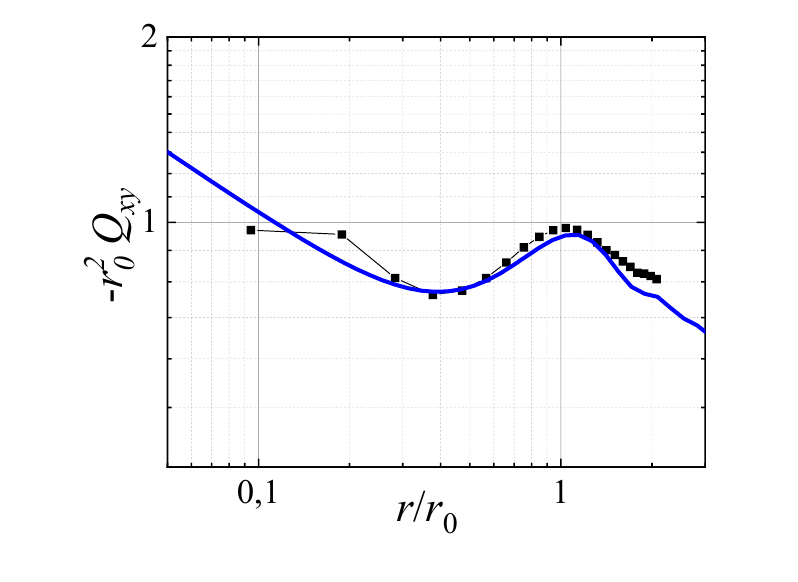}
\end{center}
\caption{The product $-r_0^2 Q_{xy}$ as a function of $r/r_0$ taken at $x=y=r/\sqrt2$ for $\sigma_R^2=0.5$, $r_0$=3.75 cm, $L_0$=20 m, where $Q_{xy}$ is the non-diagonal component of the pair correlation function of the phase gradients. The solid line is drawn using the analytic expression given in Ref. \cite{KLS25} and the filled squares represent the numerical data extracted from the simulations.}
\label{fig:Fig3}
\end{figure}

Now we pass to our numerical data for the pair correlation function of the phase gradients $Q_{\alpha\beta}$. The analytical expression for the pair correlation function in the case of the initial plane wave is given in our paper \cite{KLS25}. A comparison of the numerical data (filled squares) to the analytics (solid curve) is shown in Figs. \ref{fig:Fig2}, \ref{fig:Fig3}. We see a satisfactory agreement between the numerical data and the analytical predictions for two cases, corresponding to $\sigma_R^2=0.14$ and $\sigma_R^2=0.5$.

\begin{figure}
\begin{center}
\includegraphics[width=0.5\textwidth]{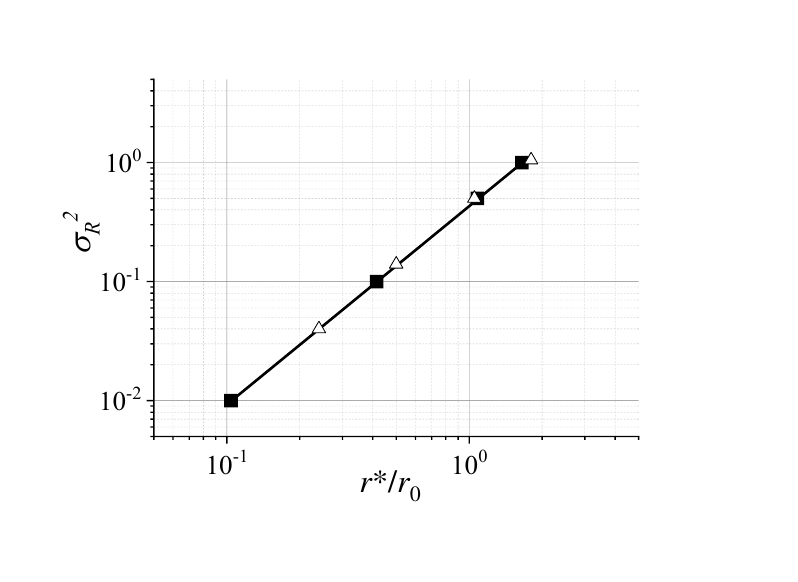}
\end{center}
\caption{A dependence of $\sigma_R^2$  on $r^\ast/r_0$, where $r^\ast$ is the separation at which the maximum of the off-diagonal component of the correlation function of the phase gradients $Q_{xy}$ is observed. The squares represent analytical predictions, the triangles are results of the numerical simulations and the solid line is determined by $\sigma_R^2=0.43(r^\ast/r_0)^{5/3}$.}
\label{fig:Fig4}
\end{figure}

One of the remarkable features of the off-diagonal component of the correlation function of the phase gradients, $Q_{\alpha\beta}$, is that it has a maximum which can be seen in Figs. \ref{fig:Fig2}, \ref{fig:Fig3}. The position of the maximum is designated as $r^\ast$. In accordance with the results of Ref. \cite{KLS25} $\sigma_R^2\propto(r^\ast/r_0)^{5/3}$. The coefficient in the proportionality law can be found numerically from the explicit analytic expressions given in Ref. \cite{KLS25}. The corresponding points are presented in Fig. \ref{fig:Fig4} by triangles, they correspond to the law $\sigma_R^2=0.43(r^\ast/r_0)^{5/3}$, see Eq. (\ref{rstar}). We see an excellent agreement between the analytical predictions and the numerical results.

\vspace{0.3cm}

Based on the numerical observations, we conclude that the zeroth-order approximation for the correlation function of phase gradients performs well at least up to $\sigma_R^2 = 0.5$.

\section{Conclusion}
\label{sec:conclusion}

Our main focus in this study is to examine numerically correlations of the radiation phase in a turbulent atmosphere, which have both fundamental and practical significance for astronomical and laser applications of adaptive optics. In Ref. \cite{BKLS24}, we proposed a new method for measuring the Fried parameter $r_0$, which plays an important role in the analysis of beams distorted by the turbulence and in determining the requirements for the spatial resolution of adaptive optical systems. The idea is to use the non-diagonal component of the correlation function of phase derivatives. The results reported in Ref. \cite{KLS25} expand the approach to arbitrary separations between the observation points.

The analytical results presented in Refs. \cite{BKLS24, KLS25} were derived in the principal order of perturbation theory for the correlation functions of the logarithm of the envelope. The investigation made in Ref. \cite{KL26} shown that the perturbation series is uniform. Therefore one expects that the results presented in Refs. \cite{BKLS24, KLS25} work well at the condition $\sigma_R^2\ll1$. Moreover, the corrections become even smaller at distances between the observation points much larger than the radius of the first Fresnel zone. Results obtained by direct numerical simulations confirm the expectations being in a good agreement with analytics.

Further, we plan to conduct experimental observations of the phase correlations to confirm our analytical and numerical findings. It should be noted that investigating the behavior of phase correlations is challenging experimentally. At low distances $r$, this is due to the finite resolution of wavefront sensors in adaptive optics systems. Additionally, when $r$ decreases and approaches the internal turbulence scale value, the expression (\ref{KolmOb}) no longer holds, and the limitation of the turbulence spectrum must be taken into account, see, e.g. Ref. \cite{Tat75}. Conversely, at high $r$, the transverse limitations of the laser beam under study or the finite size of the receiving telescope aperture affect the conditions for applicability derived for an unlimited plane wave. Nevertheless, even with these restrictions, we are able to extract detailed experimental data.

It is well known that under certain conditions, atmospheric turbulence can be non-Kolmogorov, see, e.g., Ref. \cite{Korot21}. The property is usually related to the breaking of the isotropy of turbulence, which is a complex phenomenon that requires a special investigation that lies outside the scope of this work. Note, that measurements of the off-diagonal component $Q_{xy}$ of the correlation function of the phase derivatives could provide valuable information about such non-Kolmogorov states of turbulence.

\acknowledgements

This work was supported by the scientific program of the National Center for Physics and Mathematics of RF, Section 4, stage 2026–2028.

\section*{Disclosures}

The authors declare no conflicts of interest.

\end{document}